# Anomalous size dependence of the coercivity of nanopatterned CrGeTe$_3$

*Avia Noah*[*,1,2,5], *Nofar Fridman*[1,2], *Yishay Zur*[1,2], *Maya Klang*[1], *Edwin Herrera*[3], *Jose Antonio Moreno*[3], *Martin E. Huber*[4], *Hermann Suderow*[3], *Hadar Steinberg*[1,2], *Oded Millo*[1,2], *and Yonathan Anahory**[*,1,2]

[1]The Racah Institute of Physics, The Hebrew University, Jerusalem, 9190401, Israel
[2]Center for Nanoscience and Nanotechnology, The Hebrew University of Jerusalem, Jerusalem, 91904, Israel
[3]Laboratorio de Bajas Temperaturas, Unidad Asociada UAM/CSIC, Departamento de Física de la Materia Condensada, Instituto Nicolás Cabrera and Condensed Matter Physics Center (IFIMAC), Universidad Autónoma de Madrid, E-28049 Madrid, Spain
[4]Departments of Physics and Electrical Engineering, University of Colorado Denver, Denver, CO 80217, USA
[5]Faculty of Engineering, Ruppin Academic Center, Emek-Hefer, 40250 Monash, Israel

Email: avia.noah@mail.huji.ac.il, yonathan.anahory@mail.huji.ac.il

**Abstract:**

The coercivity of single-domain magnetic nanoparticles typically decreases with the nanoparticle size and reaches zero when thermal fluctuations overcome the magnetic anisotropy. Here, we used SQUID-on-tip microscopy to investigate the coercivity of square-shaped CrGeTe$_3$ nanoislands with a wide range of sizes and width-to-thickness aspect ratios. The results reveal an anomalous size-dependent coercivity, with smaller islands exhibiting higher coercivity. The nonconventional scaling of the coercivity in CrGeTe$_3$ nanoislands was found to be inversely proportional to the island width and thickness ($1/wd$). This scaling implies that the nanoisland magnetic anisotropy is proportional to the perimeter rather than the volume, suggesting a magnetic edge state. In addition, we observe that 1600 nm wide islands display multi-domain structures with zero net remnant field, corresponding to the magnetic properties of pristine CrGeTe$_3$ flakes. Our findings highlight the significant influence of edge states on the magnetic properties of CrGeTe$_3$ and deepen our understanding of low-dimensional magnetic systems.

**Introduction**

The physics of magnetic nanoparticles has been studied thoroughly in the last few decades[1–3]. In particular, numerous studies have focused on measuring the size-dependence of the coercive field [4–7]. It is well understood that a nanoparticle smaller than some characteristic size $D_s$, becomes a single domain. The absence of domain walls typically increases the nucleation energy for a magnetic domain, thereby raising the magnetic coercivity with respect to a multi-domain particle with $D > D_s$ [4–8] (Figure 1a, blue curve). The magnetization of single-domain particles with uniaxial anisotropy is modeled as a macrospin, which is a two-level system where the energy barrier is determined by the magnetic anisotropy, which is proportional to the volume [4–8]. Due to finite-temperature fluctuations, a lower anisotropy barrier results in a lower coercivity. For sufficiently small nanoparticles, the anisotropy barrier becomes comparable to the thermal fluctuations, and the particle is in the superparamagnetic state[9] (SP, Figure 1a, left-most regime, where $H_c = 0$).

The discovery of two-dimensional (2D) van der Waals (vdW) materials with long-range magnetic order has opened a fascinating new area of magnetic materials [10–13]. The magnetic properties of these materials often differ from those of their bulk counterparts, and are thickness-dependent, thereby affording unprecedented control over their magnetism[14–17]. Experimental evidence indicates that confinement causes a transition from soft to hard ferromagnetism in $Fe_3GeTe_2$[18], $CrSiTe_3$[19], $CrGeTe_3$[20], and $CrI$[21]. Specifically, $CrGeTe_3$ (CGT) films with a thickness $d < 10$ nm exhibit a net magnetization at zero applied magnetic field[20,22]. In contrast, the interior of thicker flakes ($d > 10$ nm) has zero net magnetization at zero applied field, with hard ferromagnetism appearing only at the sample edge[20]. More recently, artificial edges fabricated by $Ga^+$ focused ion beam (FIB) etching have been shown to exhibit hard ferromagnetic (FM) properties like those of cleaved flakes, which enables direct writing of magnetic nanowires[23]. The presence of edges in narrow quasi-1D structures defined with a FIB can transform the interior region into a hard magnet. These findings raise the question of the potential influence of edges on the coercivity of CGT nanoparticles.

Here, we employed SQUID-on-tip (SOT) microscopy[24,25] at 4.2 K to measure the coercivity of FIB-patterned square-shaped CGT nanoislands with a range of island dimensions and width-to-thickness aspect ratios. Our results reveal an anomalous size dependence of the coercive field of the islands. The smallest nanoislands, with $w \times w \times d = 150 \times 150 \times 60$ nm³ ($w$ being the width), exhibit larger coercivity than the larger single-domain nanoislands with size $w \times w \times d = 600 \times 600 \times 60$ nm³ (illustrated schematically by the red curve in Figure 1a). This outcome is unexpected because coercivity usually diminishes with decreasing particle volume [4–8]. Finally, we demonstrate that islands with dimensions $1600 \times 1600 \times 60$ nm³ stabilize multi-domain structures at zero applied field and exhibit magnetic properties similar to a large pristine exfoliated CGT flake with the same thickness.

**Results**

CGT flakes were exfoliated on top of a SiO$_2$-coated Si wafer. We etched vertical and horizontal lines using a 30 keV Ga$^+$ FIB, resulting in an array of square-shaped magnetic islands. The synthesis and fabrication details are presented in Supplementary Note 1. The effect of island geometry was examined by varying the effective crystal thickness from $d = 25$ to 70 nm and the width from $w = 150$ to 1600 nm. The island dimensions was measured by cross-sectional STEM, see Supplementary Note 2 for details. The number of islands in each array varied between 81 and 121 for islands with $w < 1600$ nm (see Supplementary Table 1 for details). Probing the properties of a large number of island allow us to reduce the statistical uncertainty (See Supplementary Note 3). The out-of-plane component of the magnetic field, $B_z(x,y)$, emanating from the array, was imaged at 4.2 K using a scanning SOT microscope[24,25], as presented schematically in Figure 1a (see also Supplementary Note 4). Figures 1b-d represent magnetic images of islands acquired near the coercive field of the respective array, where the black/white color-coded area indicates a magnetic moment pointing downwards/upwards. The array parameters of the islands are: thickness, $d = 60$ nm for all, and widths, $w = 150,\ 600$, and 1600 nm in Figures 1b-d, respectively. The results indicate that small islands ($w \leq 600$ nm, Figs. 1b, c), are single-domain, while larger islands with $w = 1600$ nm, exhibit fragmentation into multiple magnetic domains (Fig. 1d).

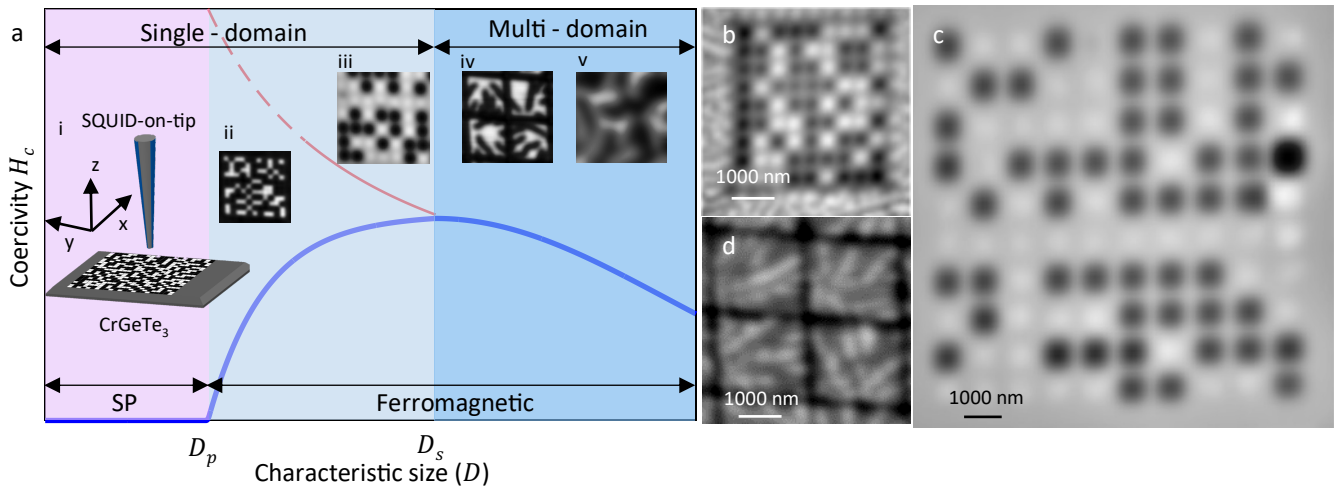

**Figure 1. SOT images of the island arrays patterned in CrGeTe$_3$ via FIB**. (**a**) (blue curve) A schematic representation of the typical coercive field $H_c$ dependence on the nanoparticle characteristic size $D$. (red solid curve) Same as blue curve but based on the experimental results for CrGeTe$_3$. (red dashed curve) Extrapolation based on our results. **Insets:** (**i**) Schematic illustration of the SQUID-on-tip (SOT) measurement. (ii-v) Typical $B_z(x,y)$ images of CGT island with distinct dimensions (**ii-iii**) Single magnetic domain array with characteristic size smaller than $D_s$. (**iv**) $B_z(x,y)$ images of four multi-domain CGT islands with width $w = 1600$ nm $> D_s$. (**v**) Unpatterned CGT flake with $w \sim 10$ μm and $d = 50$ nm. The image size is $5 \times 5$ **ii-iii**, $4.2 \times 4.2$ **iv** and $3 \times 3$ μm$^2$ **v**. (**b-d**) $B_z(x,y)$ images acquired near the coercive field of the corresponding array. The island dimensions are $d = 60$ nm for all arrays, with widths, $w = 150$ nm **b**, 600 nm **c**, and 1600 nm **d**. Imaging parameters: (**b**) $\mu_0 H_z = 70$ mT, area scan $4.1 \times 4.1$ μm$^2$, pixel size 32 nm, (**c**) $\mu_0 H_z = 20$ mT, area scan $11 \times 11$ μm$^2$, pixel size 115 nm, and (**d**) $\mu_0 H_z = 100$ mT, area scan $4.2 \times 4.2$ μm$^2$, pixel size 30 nm. The scale bar is 1000 nm in b,c,d. The black to white color scale represents lower and higher magnetic fields, respectively.

We characterize the magnetic response of the arrays to an applied out-of-plane magnetic field $H_z$, by counting the number of islands pointing in a given direction. Figure 2a plots the resulting normalized magnetization curves, $M(H_z)/M_{tot}$, where $M_{tot} = N|m_i|$; $N$ is the number of islands, and $m_i$ is the magnetic moment per island. $m_i$ is estimated by using the volumetric spin density ($\approx$ 3 μ$_B$/Cr) that was found to be constant down to a few layers[26]. This value is consistent with our bulk magnetic measurements (Supplementary Note 2).

The hysteretic curves of all the arrays display a smooth magnetization reversal. The range of fields over which magnetization reversal occurs, the transition width, $\Delta H_z$, was found to be similar for all measured arrays $\Delta H_z = H_l - H_f = 73 \pm 7$ mT, where $H_f$ ($H_l$) is the field at which the first (last) island reverses its magnetization. The microscopic mechanism responsible for the island variability remains unknown. Here, the island variability is treated as an additional uncertainty on the individual island coercive field (see Supplementary Note 3).

The coercive field of the array $H_c^a$ is reached when $M(H_c^a) = 0$, or when the magnetization of half of the islands point in a given direction. Therefore, $H_c^a$ is the median value of the field at which a single-island reverses its magnetization, defined as $\widetilde{H_c^i}$. Notably, $\widetilde{H_c^i}$ varies significantly with the island geometry, and ranges from 15 to 100 mT (black dots in Figure 2a). In the absence of a stable magnetic domain wall, the magnetic saturation field is $H_c^i = 2K/m_i$, where $K$ is the island magnetic anisotropy. Therefore, measuring $H_c^a$ allows us to determine the median single-island magnetic anisotropy $K$. For larger particles, $K \gg k_b T$, $H_c^i$ is expected to be constant given that $K$ and $m_i$ are proportional to the island volume. However, for smaller particles, $H_c^i$ is expected to decrease and reach zero when $K \sim k_b T$. In striking contrast, we observe the opposite trend, and the results indicate that islands with smaller volumes tend to exhibit larger values of $\widetilde{H_c^i}$ (Figure 2a).

Notably, $\widetilde{H_c^i}$ is not fully determined by the island's volume, given that islands with comparable volumes, $V_1 = 1.4 \pm 0.1 \times 10^6$ nm³ and $V_3 = 1.4 \pm 0.1 \times 10^6$ nm³, but different aspect ratios, have significantly different values, with $\widetilde{H_c^i} = 99 \pm 7$ and $70 \pm 7$ mT, respectively (Fig. 2a and table 1). Distinct values of $\widetilde{H_c^i}$ are measured in arrays with the same thickness but different widths. Arrays $V_3$ and $V_7$, with $d = 60 \pm 2$ nm and widths $w = 150 \pm 5$ and $600 \pm 5$ nm, have values of $\widetilde{H_c^i} = 70 \pm 7$ and $20 \pm 7$ mT, respectively. Finally, arrays $V_6$ and $V_7$, with comparable widths, $w = 550 \pm 5$ and $600 \pm 5$ nm, not only differ in values of $\widetilde{H_c^i} = 60 \pm 7$ and $20 \pm 7$ mT, respectively, but also undergo a transition from ideal hard FM ($|M(H_z = 0)/M_{tot}| = 1$) to a softer FM ($|M(H_z = 0)/M_{tot}| < 1$). Taking the influence of the demagnetizing factor into account, supplementary Figure 1 presents a plot of $H_c^i$ vs. the shape anisotropy of each island[27], and lacks evidence of any trend in the observed data (see values in Table 1).

Examining arrays with the same thickness, $d = 35$ nm, and varying widths, $w = 200, 220,$ and 240 nm ($V_1$, $V_2$ and $V_4$), revealed a monotonic trend in $H_c$ (see Table 1 and Figure 2a). Previous results revealed that the magnetism at the edge of thick flakes with $d > 10$ nm, has finite coercivity in contrast to the sample interior, which has no net remnant field[3]. Assuming that the anisotropy barrier is governed by a one-dimensional magnetic edge state, we find that $K$ is proportional to $w$ (the island perimeter) and not to the volume, as is commonly observed [4–8]. As a result, we find that $\widetilde{H_c^i} = 2K/m_i \propto w/V = 1/wd$, where $V = w^2 d$. Figure 2b reveals the linear relation between the measured $\widetilde{H_c^i}$ and $1/wd$. This experimental evidence thus demonstrates that the magnetic properties of the island are governed by a one-dimensional magnetic edge state.

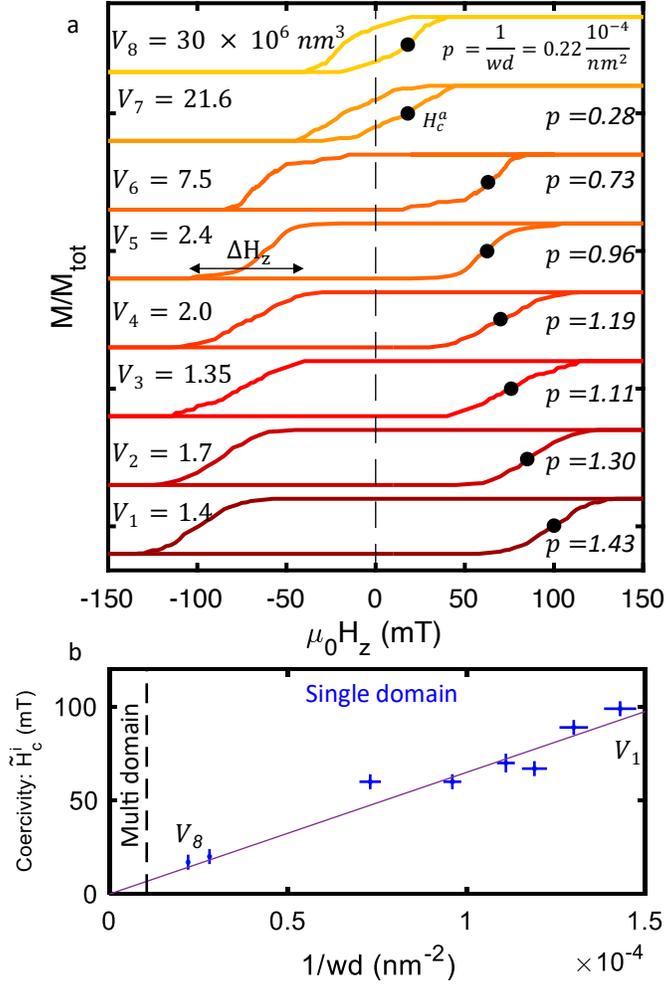

**Figure 2. Field evolution of island arrays in CrGeTe₃.** (**a**) Hysteresis curves drawn from $B_z(x,y)$ measured on arrays with volumes $V$ ranging between $1.35 \times 10^6$ and $30 \times 10^6$ nm³. The array's coercive field $H_c^a$ is marked with black dots. The hysteresis curves were measured by ramping the field in one direction and were symmetrized to obtain the second branch of the $M(H_z)/M_{tot}$ curve (see Supplementary Note 3). Curves were shifted vertically for clarity. (**b**) The median island coercive field, $H_c^i$ ($\widetilde{H_c^i}$), as a function of the parameter $w/V = 1/wd$. Several SOT images are shown in Supplementary Figure 2.

| Array # | Effective size $w \times w \times d$ (nm³) | Volume ($10^6$ nm³) | Width-to-volume ratio $p = \dfrac{w}{V} = \dfrac{1}{wd}$ ($10^{-4}$ nm$^{-2}$) | Shape anisotropy ($10^{-6}$ eV nm$^{-3}$) | Island Magnetization $m_i = \dfrac{3\mu_b V}{V_{cell}}$ (eV T$^{-1}$) | Median island Coercivity $\widetilde{H_c^i}$ (mT) | Transition width $\Delta H_z = H_l - H_f$ (mT) | Median island anisotropy $K = \dfrac{H_c^i M}{2}$ (eV) |
|---|---|---|---|---|---|---|---|---|
| 1 | 200 × 200 × 35 | 1.4 ± 0.1 | 1.43 ± 0.09 | 5.0 ± 0.4 | 290 ± 20 | 99 ± 4 | 76 ± 6 | 14 ± 2 |
| 2 | 220 × 220 × 35 | 1.7 ± 0.1 | 1.30 ± 0.08 | 5.3 ± 0.4 | 360 ± 30 | 89 ± 4 | 72 ± 6 | 15 ± 2 |
| 3 | 150 × 150 × 60 | 1.4 ± 0.1 | 1.11 ± 0.05 | 2.8 ± 0.3 | 280 ± 20 | 70 ± 5 | 82 ± 7 | 10 ± 2 |
| 4 | 240 × 240 × 35 | 2.0 ± 0.1 | 1.19 ± 0.07 | 5.5 ± 0.4 | 420 ± 30 | 67 ± 4 | 80 ± 7 | 14 ± 2 |
| 5 | 230 × 230 × 45 | 2.4 ± 0.1 | 0.96 ± 0.05 | 4.8 ± 0.3 | 500 ± 30 | 60 ± 4 | 75 ± 6 | 15 ± 2 |
| 6 | 550 × 550 × 25 | 7.5 ± 0.6 | 0.73 ± 0.06 | 7.3 ± 0.2 | 1600 ± 100 | 60 ± 4 | 77 ± 6 | 47 ± 8 |
| 7 | 600 × 600 × 60 | 21.6 ± 0.8 | 0.28 ± 0.01 | 6.2 ± 0.1 | 4500 ± 200 | 20 ± 4 | 65 ± 6 | 50 ± 20 |
| 8 | 650 × 650 × 70 | 30 ± 1 | 0.220 ± 0.007 | 6.1 ± 0.1 | 6200 ± 200 | 17 ± 4 | 61 ± 6 | 50 ± 20 |

**Table 1. A summary of the islands' parameters and results presented in Figure 2.** The uncertainty on the dimension is ±5 nm for the width $w$ and ±2 nm for the thickness $d$.

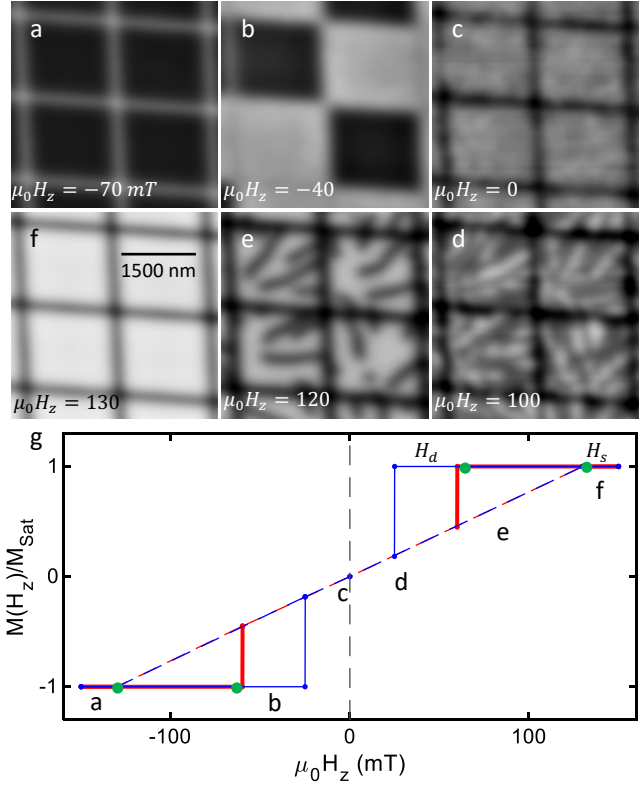

**Figure 3. Magnetic field response of the CrGeTe$_3$ with $w \times w \times d = 1600 \times 1600 \times 60$ nm³ islands.** (a-f) Sequence of SQUID-on-tip $B_z(x,y)$ images at distinct values of an applied out-of-plane field $\mu_0 H_z$. (g) Illustrated hysteresis curves drawn from $B_z(x,y)$ measured on array that comprise 16 islands of volume $V = 1600 \times 1600 \times 60$ nm³. The red/blue curve correspond to the island with largest/smallest demagnetization field $H_d$. The green dots represent $H_d$ and saturation field $H_s$ measured for pristine 60 nm thick CrGeTe$_3$ flake, extracted from Ref. [20]. The dashed lines represent the fields at which the islands are in the multi-domain state. Imaging parameters: $\mu_0 H_z = $ -70 **a**, -40 **b**, 0 **c**, 100 **d**, 120 **e**, and 130 **f** mT. Area scan 4.2 × 4.2 µm², pixel size 30 nm. The black to white color scale represents lower and higher magnetic field, respectively. The color scale is 16, 24, 3, 3, 9, and 18 mT for a-f, respectively.

For larger islands with $w = 1600$ nm and $d = 60$ nm, a multi-domain state is observed during the magnetization reversal (Figure 1d). Figure 3 presents a sequence of $B_z(x, y)$ SOT images of four islands between negative to positive saturation. After excursion of $\mu_0 H_z = -200$ mT, all islands hold their magnetization up to $\mu_0 H_z = -70$ mT (Figure 3a uniform black color code). In contrast to smaller islands, here the islands' magnetization breaks into a multi-domain state at demagnetization fields in the range $\mu_0 H_d = -65$ to $-20$ mT. Figure 3b acquired at $\mu_0 H_z = -40$ mT depicts two islands at saturation magnetization (black) and two islands in the multi-domain state (gray). The magnetic domains are smaller than the tip diameter (~ 150 nm), resulting in a magnetic contrast of 1 mT. At $\mu_0 H_z = 0$ mT, all the islands are in the multi-domain states where hard ferromagnetism is observed on the edges, resulting in a negative stray magnetic field as reported previously for patterned CGT[23] (Supplementary Figure 3). Increasing the out-of-plane field caused the magnetic domains with a magnetic moment parallel to the field to grow at the expense of domains with antiparallel magnetic moments. Figures 3d–e demonstrate that the last domains to reverse their magnetization are long and thin, and tend to touch the edge of the island. At the saturation field of $\mu_0 H_z = 130$ mT, the islands' magnetization is saturated in the positive direction and parallel with the external magnetic field (Figure 3f).

Figure 3g illustrated two sketched magnetization curves of distinct individual island drawn from SOT images of sixteen islands. Each individual island exhibit slightly different properties. In particular, $H_d$ varied from $H_d = -65$ to $-20$ mT and the corresponding two extreme cases are plotted in red and blue curves, respectively (see images taken at such field in Supplementary figure 4). The dashed lines represent the fields at which the islands are in the multi-domain state up to a saturation field $H_s = 130$ mT, which shows significantly less variability than $H_d$. The reason for the better reproducibility of $H_s$ over $H_d$ is not understood and will require more investigation. In the multi-domain state, previous measurements showed no magnetic hysteresis[20]. We compare the illustrated hysteresis loops of individual islands with the data measured previously on a pristine (exfoliated) CGT flake[20]. Notably, the saturation field $H_s = 130$ mT, and a demagnetization field $H_d$ of -66 mT (represented by the green dots), are close to the values seen in the red curve. We conclude that for this array, the islands hysteresis loops are comparable to that of the pristine CGT flake (same value of $H_s$ and $H_d$ values being in the same range). This contrasts with arrays with smaller island volumes that exhibit hard ferromagnetism.

**Discussion**

This study was designed to investigate the magnetic properties of CGT nanoislands as a function of size and aspect ratio. The results reveal a transition from hard ferromagnetic single-domain islands to multi-domain and zero remnant field islands. For $d = 60$ nm, the transition occurs between $w = 650$ and $1600$ nm. A similar transition was previously observed in FIB-patterned CGT stripes with a length of 10 μm and $d = 50$ nm[23]. In the case of the stripe geometry, the transition occurred between $w = 270$ and $400$ nm, which is significantly smaller than the currently observed values for islands. The larger $w$ length scale range over which CGT remains a hard ferromagnet in the present study is consistent with the larger (factor two) perimeter-to-volume ratio in an island compared to a stripe of length much larger than width.

Another manifestation of the island edge is the unique dependence of the island saturation field on the island geometry, namely, $H_c^i \propto 1/wd$. This unusual dependence can be explained by assuming the magnetic anisotropy, $K$, scales with the island perimeter rather than the volume. The microscopic mechanism causing this edge state is currently unknown. Several mechanisms were considered in the past[20,23]. One of them is related to the in-plane dangling bonds. If such a mechanism would be dominant, one should find magnetism also at step-edges between two terraces. Previous work showed the absence of magnetism at such step-edges (see Figure S9 in Ref. 20) and suggests that this scenario is less probable. Moreover, magnetic edges were found at the edges of cleaved samples exposed to air, encapsulated[20] and amorphized[23], reinforcing the idea that the effect of in-plane dangling bound is negligible. Gallium contamination was also considered as a potential mechanism. However, magnetic edges were found in samples that were never exposed to the Ga beam. Moreover, we did not observe any sign of higher gallium concentration at the edge of the crystalline part of the island (see Supplementary Figure 6), ruling out the potential role of gallium contamination. It is plausible that some strain appears at low temperatures at the interface between the amorphous and crystalline CGT regions. However, given that such magnetic edge state was found in cleaved samples, where such interface does not exist, tends to rule out this scenario. Another plausible mechanism is the presence of strain at the sample edges[28,29]. Recently, similar edge state in another material is thought to be caused by the Stoner mechanism[30]. Notably, applying this model to CGT is not straightforward since CGT is insulating and further investigation will be necessary to confirm the presence of such a state in CGT.

Another consequence of the finding that the magnetic anisotropy is proportional to $w$ is that $H_c^i$ will not grow indefinitely with reducing size; it is rather expected to reach zero at the blocking temperature $T_B = K/25k_b$[8]. In the range of sizes investigated here, we find $K = 0.078 \times 10^{-9}\ w$. By setting $T_B = 4$ K, we can estimate that the smallest $w$ that would result in a finite $H_c^i$ is $w \sim 0.1$ nm. This is a non-physical dimension given that it is smaller than the lattice constant. We can therefore conclude that our model ($K \propto w$) must break down at a length larger than 0.1 nm. Previous results indicated that the magnetism of two-dimensional flakes becomes undetectable below $\sim 7$ layers[26]. Although the out-of-plane coupling is usually weaker in magnetic van der Waals materials, we expect that the nanoislands could be scaled down to such dimensions. In the current study we were not able to reach this regime, since the smallest island achievable with Ga$^+$ based FIB is of width $w \sim 100$ nm. Other techniques, such as He based FIB, have a higher resolution that could allow us to investigate smaller sizes.

To conclude, our results demonstrate that we can adjust the local magnetic properties of CGT by controlling the dimensions, here achieved by using Ga$^+$ FIB fabrication of square-shaped nanoislands. We report an anomalous size-dependence of island coercivity, which is inversely proportional to the width and thickness. In addition, we observe the transition between single and multi-domain above a critical width. Notably, controlling ferromagnetic order in vdW heterostructures may play a substantial role in spintronic devices[31–34] and in the study of proximity-induced phenomena[35,36].

**Conflict of interest**

There are no conflicts to declare.

**Acknowledgements:**

We would like to thank O. Agam, M. Kläui, and A. Capua for fruitful discussions. We thank A. Vakahi and S. Remennik for technical support and J. L. Martínez for his support in performing the bulk magnetic characterization. This work was supported by the European Research Council (ERC) Foundation grant No. 802952 and the Israel Science Foundation (ISF) Grant No. 645/23. The international collaboration on this work was fostered by the EU-COST Action CA21144 (Superqmap). H. Steinberg acknowledges funding provided by the DFG Priority program grant 443404566 and Israel Science Foundation (ISF) grant 164/23. O. Millo is grateful for support from ISF grant no. 576/21 and the Harry de Jur Chair in Applied Science. H. Suderow and E. Herrera acknowledge the Spanish Research State Agency (PID2020-114071RB-I00, CEX2023001316-M, TED2021-130546B-I00), by the Comunidad de Madrid through program NANOMAGCOST-CM (Program No.S2018/NMT-4321) .

**Author Contributions:**

Y.A., and A.N. conceived the experiment.
E.H. and H.S. synthesized the CGT crystals.
J.M. characterized the CGT crystals.
A.N., Y.Z., and N.F. carried out the scanning SOT measurements.
Y.A., M.K., H. Steinberg, and A.N. fabricated the CGT devices.
A.N. characterized the CGT devices.
A.N. analyzed the data.
Y.A., and A.N. constructed the scanning SOT microscope.
M.E.H. developed the SOT readout system.
A.N., O.M., and Y.A. wrote the paper with contributions from all authors.
Notes: The authors declare no competing financial interest.

**References**

1. Ge, J., Hu, Y., Biasini, M., Beyermann, W. P. & Yin, Y. Superparamagnetic Magnetite Colloidal Nanocrystal Clusters. *Angewandte Chemie International Edition* **46**, 4342–4345 (2007).
2. Duan, M., Shapter, J. G., Qi, W., Yang, S. & Gao, G. Recent progress in magnetic nanoparticles: synthesis, properties, and applications. *Nanotechnology* **29**, 452001 (2018).
3. Ma, Z., Mohapatra, J., Wei, K., Liu, J. P. & Sun, S. Magnetic Nanoparticles: Synthesis, Anisotropy, and Applications. *Chem Rev* **123**, 3904–3943 (2023).
4. Sung Lee, J., Myung Cha, J., Young Yoon, H., Lee, J.-K. & Keun Kim, Y. Magnetic multi-granule nanoclusters: A model system that exhibits universal size effect of magnetic coercivity. *Sci Rep* **5**, 12135 (2015).
5. Kneller, E. F. & Luborsky, F. E. Particle Size Dependence of Coercivity and Remanence of Single-Domain Particles. *J Appl Phys* **34**, 656–658 (1963).
6. Sharrock, M. P. Time-dependent magnetic phenomena and particle-size effects in recording media. *IEEE Trans Magn* **26**, 193–197 (1990).
7. Lisjak, D. & Mertelj, A. Anisotropic magnetic nanoparticles: A review of their properties, syntheses and potential applications. *Prog Mater Sci* **95**, 286–328 (2018).
8. Cullity, B. D. & Graham, C. D. *Introduction to Magnetic Materials*. Introduction to Magnetic Materials (Wiley, 2008). doi:10.1002/9780470386323.
9. Bean, C. P. & Livingston, J. D. Superparamagnetism. *J Appl Phys* **30**, S120–S129 (1959).
10. Yang, S., Zhang, T. & Jiang, C. van der Waals Magnets: Material Family, Detection and Modulation of Magnetism, and Perspective in Spintronics. *Advanced Science* **8**, 2002488 (2021).
11. Deng, Y. *et al.* Layer-Number-Dependent Magnetism and Anomalous Hall Effect in van der Waals Ferromagnet $Fe_5GeTe_2$. *Nano Lett* **22**, 9839–9846 (2022).
12. Zhang, W., Wong, P. K. J., Zhu, R. & Wee, A. T. S. Van der Waals magnets: Wonder building blocks for two-dimensional spintronics? *InfoMat* **1**, 479–495 (2019).
13. Jin, C. & Kou, L. Two-dimensional non-van der Waals magnetic layers: functional materials for potential device applications. *J Phys D Appl Phys* **54**, 413001 (2021).
14. Thiel, L. *et al.* Probing magnetism in 2D materials at the nanoscale with single-spin microscopy. *Science (1979)* **364**, 973–976 (2019).
15. Wang, Q. H. *et al.* The Magnetic Genome of Two-Dimensional van der Waals Materials. *ACS Nano* **16**, 6960–7079 (2022).
16. Song, T. *et al.* Direct visualization of magnetic domains and moiré magnetism in twisted 2D magnets. *Science (1979)* **374**, 1140–1144 (2021).
17. Niu, B. *et al.* Coexistence of Magnetic Orders in Two-Dimensional Magnet CrI3. *Nano Lett* **20**, 553–558 (2020).
18. Tan, C. *et al.* Hard magnetic properties in nanoflake van der Waals Fe3GeTe2. *Nat Commun* **9**, 1554 (2018).
19. Zhang, C. *et al.* Hard ferromagnetic behavior in atomically thin CrSiTe3 flakes. *Nanoscale* **14**, 5851–5858 (2022).

20. Noah, A. *et al.* Interior and Edge Magnetization in Thin Exfoliated CrGeTe3 Films. *Nano Lett* **22**, 3165–3172 (2022).
21. Huang, B. *et al.* Layer-dependent ferromagnetism in a van der Waals crystal down to the monolayer limit. *Nature* **546**, 270–273 (2017).
22. Gong, C. *et al.* Discovery of intrinsic ferromagnetism in two-dimensional van der Waals crystals. *Nature* **546**, 265–269 (2017).
23. Noah, A. *et al.* Nano-Patterned Magnetic Edges in CrGeTe3 for Quasi 1-D Spintronic Devices. *ACS Appl Nano Mater* **6**, 8627–8634 (2023).
24. Anahory, Y. *et al.* SQUID-on-tip with single-electron spin sensitivity for high-field and ultra-low temperature nanomagnetic imaging. *Nanoscale* **12**, 3174–3182 (2020).
25. Vasyukov, D. *et al.* A scanning superconducting quantum interference device with single electron spin sensitivity. *Nat Nanotechnol* **8**, 639–644 (2013).
26. Vervelaki, A. *et al.* Visualizing thickness-dependent magnetic textures in few-layer Cr2Ge2Te6. *Commun Mater* **5**, 40 (2024).
27. Aharoni, A. Demagnetizing factors for rectangular ferromagnetic prisms. *J Appl Phys* **83**, 3432–3434 (1998).
28. Šiškins, M. *et al.* Nanomechanical probing and strain tuning of the Curie temperature in suspended Cr2Ge2Te6-based heterostructures. *NPJ 2D Mater Appl* **6**, 41 (2022).
29. O'Neill, A. *et al.* Enhanced Room Temperature Ferromagnetism in Highly Strained 2D Semiconductor Cr 2 Ge 2 Te 6. *ACS Nano* **17**, 735–742 (2023).
30. Liu, S. *et al.* Surface-induced ferromagnetism and anomalous Hall transport at Zr2S (001). *Phys Rev Mater* **7**, 024409 (2023).
31. Jungwirth, T., Marti, X., Wadley, P. & Wunderlich, J. Antiferromagnetic spintronics. *Nat Nanotechnol* **11**, 231–241 (2016).
32. Hirohata, A. *et al.* Review on spintronics: Principles and device applications. *J Magn Magn Mater* **509**, 166711 (2020).
33. Telford, E. J. *et al.* Coupling between magnetic order and charge transport in a two-dimensional magnetic semiconductor. *Nat Mater* **21**, 754–760 (2022).
34. Kurebayashi, H., Garcia, J. H., Khan, S., Sinova, J. & Roche, S. Magnetism, symmetry and spin transport in van der Waals layered systems. *Nat Rev Phys* **4**, 150–166 (2022).
35. Han, W., Kawakami, R. K., Gmitra, M. & Fabian, J. Graphene spintronics. *Nat Nanotechnol* **9**, 794–807 (2014).
36. Linder, J. & Robinson, J. W. A. Superconducting spintronics. *Nat Phys* **11**, 307–315 (2015).

# Supplementary Material

# Anomalous size dependence of the coercivity of nanopatterned CrGeTe$_3$

*Avia Noah*[1,2,5], *Nofar Fridman*[1,2], *Yishay Zur*[1,2], *Maya Klang*[1], *Edwin Herrera*[3], *Jose Antonio Moreno*[3], *Martin E. Huber*[4], *Hermann Suderow*[3], *Hadar Steinberg*[1,2], *Oded Millo*[1,2], *and Yonathan Anahory**[1,2]

[1]The Racah Institute of Physics, The Hebrew University, Jerusalem, 9190401, Israel
[2]Center for Nanoscience and Nanotechnology, The Hebrew University of Jerusalem, Jerusalem, 91904, Israel
[3]Laboratorio de Bajas Temperaturas, Unidad Asociada UAM/CSIC, Departamento de Física de la Materia Condensada, Instituto Nicolás Cabrera and Condensed Matter Physics Center (IFIMAC), Universidad Autónoma de Madrid, E-28049 Madrid, Spain
[4]Departments of Physics and Electrical Engineering, University of Colorado Denver, Denver, CO 80217, USA
[5]Faculty of Engineering, Ruppin Academic Center, Emek-Hefer, 40250 Monash, Israel

Email: avia.noah@mail.huji.ac.il, yonathan.anahory@mail.huji.ac.il

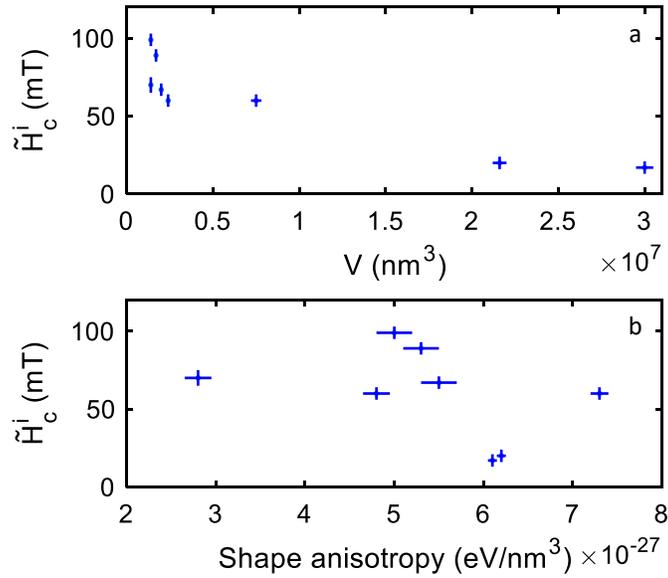

**Supplementary Figure 1 - The median island coercive field, $\widetilde{H}_c^i$.** (a) $\widetilde{H}_c^i$ as a function of island volume $V$. (b) $\widetilde{H}_c^i$ as a function of the island anisotropy.

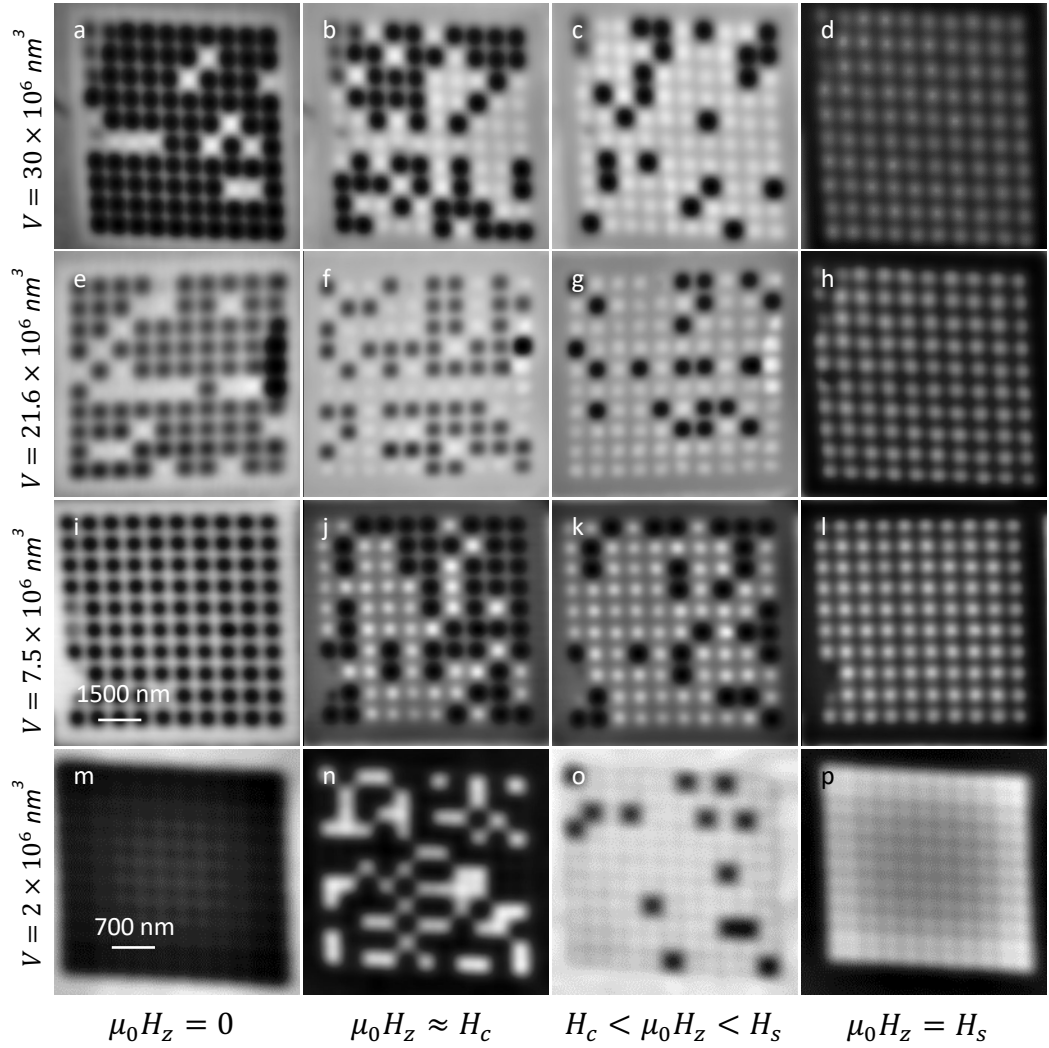

**Supplementary Figure 2 - SOT images of nano-patterned island array in CrGeTe$_3$.** (**a-p**) Sequence of magnetic images $B_z(x,y)$ of different island volumes at distinct values of applied out-of-plane field $\mu_0 H_z$. Imaging parameters: (**a-d**) $V = 30 * 10^6$ nm$^3$, scan area 11 ×11 μm$^2$, pixel size 115 nm. $\mu_0 H_z = 0$ **a**, 15 **b**, 30 **c**, 50 mT **d**. (**e-h**) $V = 22 * 10^6$ nm$^3$, scan area 11 ×11 μm$^2$, pixel size 115 nm. $\mu_0 H_z = 0$ **e**, 20 **f**, 30 **g**, 50 mT **h**. (**i-l**) $V = 7.5 * 10^6$ nm$^3$, scan area 11×11 μm$^2$, pixel size 115 nm. $\mu_0 H_z = 0$ **l**, 60 **j**, 70 **k**, 100 mT **l**. (**m-p**) $V = 2 * 10^6$ nm$^3$, scan area 5 × 5 μm$^2$, pixel size 40 nm. $\mu_0 H_z = 0$ **m**, 70 **n**, 90 **o**, 120 mT **p**. The black to white color scale represents lower to higher magnetic fields, respectively.

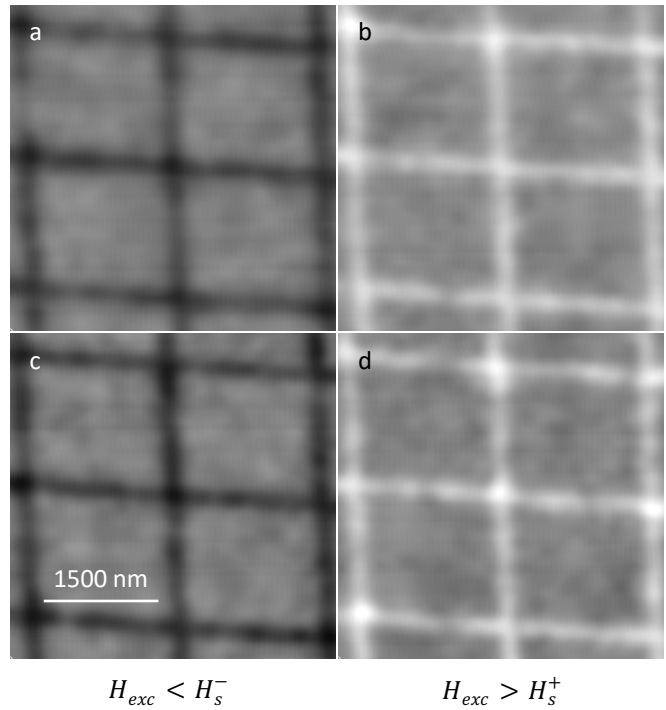

$H_{exc} < H_s^-$    $H_{exc} > H_s^+$

**Supplementary Figure 3 - SOT images of magnetic edges due to amorphization in CrGeTe$_3$ at zero field.** (a-d) Sequence of magnetic images $B_z(x,y)$ of amorphized CGT islands after distinct field excursions. (a, c) $H_{exc} < H_s^-$, (b, d) $H_{exc} > H_s^+$. The flake thickness is d = 60 nm. Imaging parameters: $\mu_0 H_z = 0$ mT, area scan 4.2 × 4.2 µm², pixel size 30 nm. The black to white color scale represents a lower and higher magnetic field, respectively.

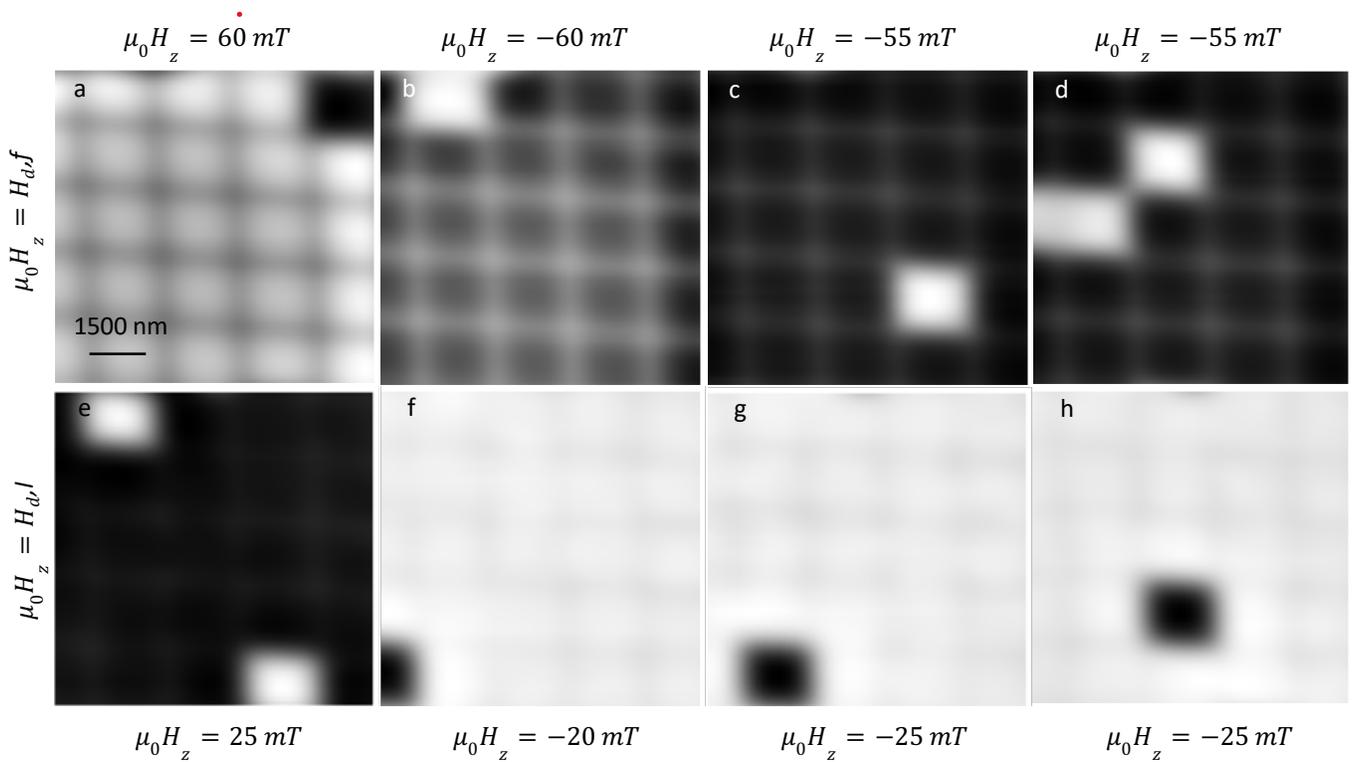

**Supplementary Figure 4 – SOT images of the first and last island to demagnetize** (a-h) Magnetic images $B_z(x,y)$ acquired with the SOT that shows the first **a-d** and last **e-h** to demagnetize in different set of measurements. The field at which the first and last particle demagnetizes is labeled $H_{d,f}$ and $H_{d,l}$, respectively. The black to white color scale represents lower and higher magnetic field, respectively. Each image is 8.5 × 8.5 µm² and comprise 64 × 64 pixels.

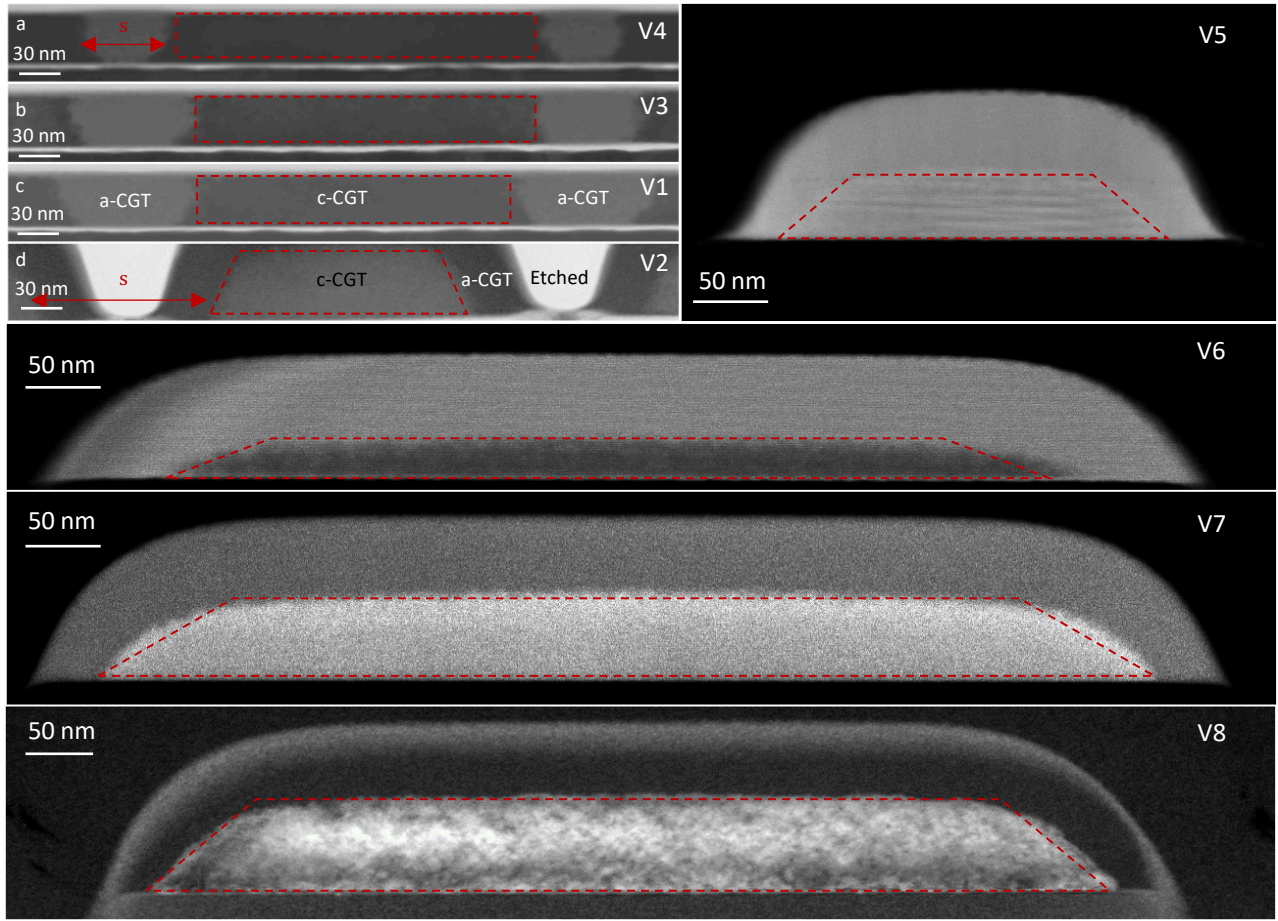

**Supplementary Figure 5 - High angle annular dark-HAADF image of each volume the arrays.** The area inside the dark red dotted line trapezoid is the region where magnetic crystalline CrGeTe$_3$ is found. Above the crystalline region, an amorphous region is observed for volumes $V_2$, $V_5$, $V_6$, $V_7$ and $V_8$.

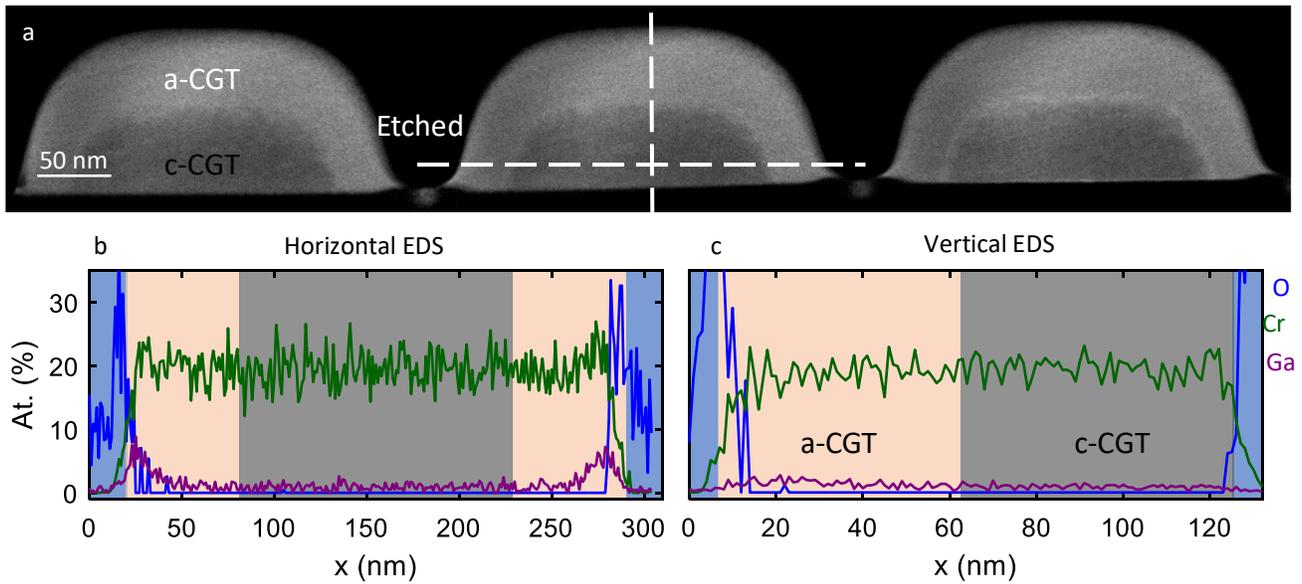

**Supplementary Figure 6 – EDS measurements of CrGeTe₃ island. (a)** High-angle annular dark field (HAADF) image of the $V_3$ array ($150 \times 150 \times 60$ nm³). **(b)** Energy-Dispersive X-ray Spectroscopy (EDS) cross-sections, showing the relative amount of Cr, O, and Ga in a cross section of the island.

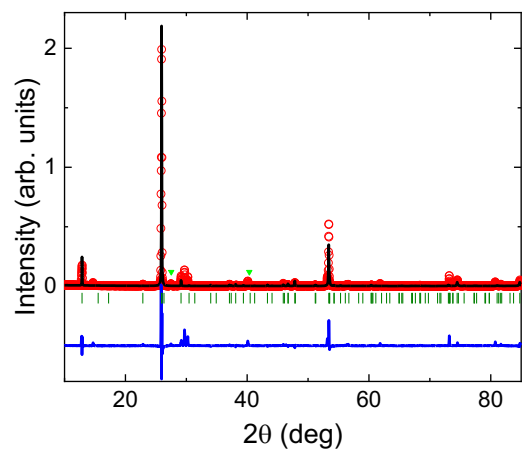

**Supplementary Figure 7 - Powder diffraction pattern of CrGeTe₃.** Red symbols are the experimental points. The black line is the best fit to CrGeTe₃ diffraction pattern with refined lattice parameters a = 0.6809 nm, b = 0.6809 nm, c = 2.0444 nm. Residuals are given by the blue line. The vertical green strikes represent the position in 2θ scale of the reflections from the CrGeTe₃ (space group R-3h (148)). Green triangles identify the effects of Te flux.

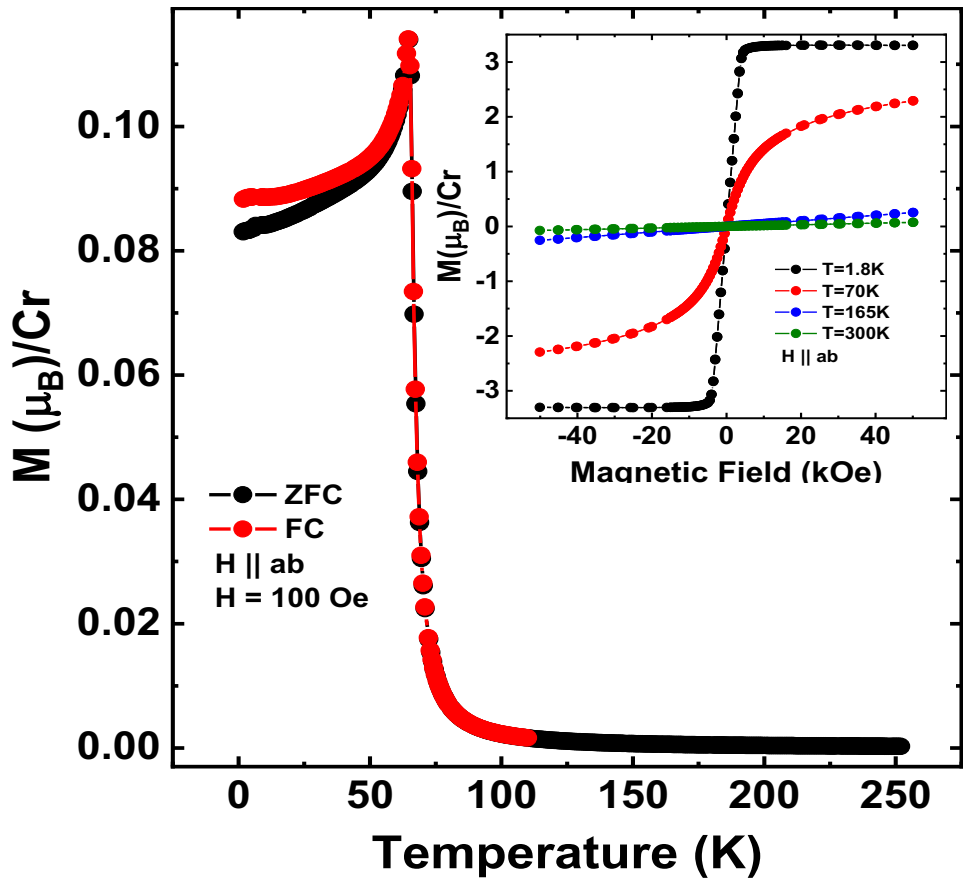

**Supplementary Figure 8 – Bulk M(H) measurements.** Magnetization as a function of the temperature for ZFC (black symbols) and FC (red symbols) for a magnetic field of 100 Oe applied parallel to the ab-plane. Inset shows Magnetization as a function of the magnetic field for different temperatures.

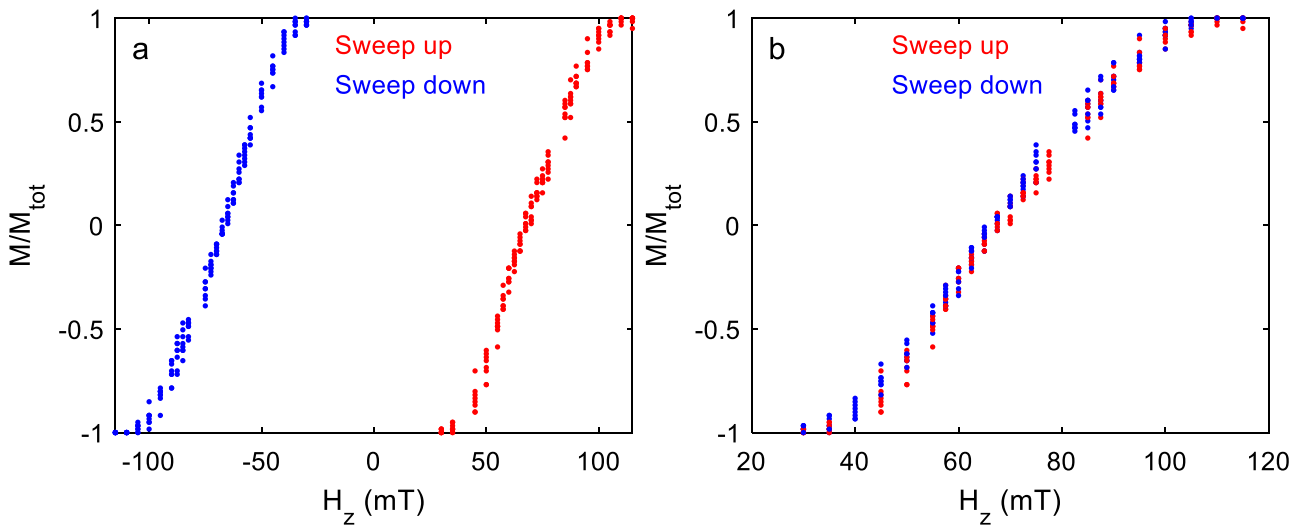

**Supplementary Figure 9 - – Measurement reproducibility..** (a) Hysteresis curves drawn from $B_z(x, y)$ images measured for $V_4$ over 8 complete hysteresis loops. (b) Same data as in **a** but showing $M(H)$ for the sweep up (blue dots) and $-M(-H)$ for the sweep down (red dots).

| Array name | Number of measured loops | Steps in field (mT) $\delta H_z$ | Number of Island in the array $N$ | Inter-island Spacing (nm) |
|---|---|---|---|---|
| $V_1$ | 8 | 2.5 | $11 \times 11$ | 100 |
| $V_2$ | 7 | 2.5 | $11 \times 11$ | 80 |
| $V_3$ | 3 | 2.0 | $9 \times 9$ | 200 |
| $V_4$ | 8 | 2.5 | $11 \times 11$ | 60 |
| $V_5$ | 3 | 2.5 | $10 \times 10$ | 200 |
| $V_6$ | 5 | 2.0 | $10 \times 10$ | 200 |
| $V_7$ | 4 | 2.0 | $10 \times 10$ | 200 |
| $V_8$ | 6 | 2.0 | $10 \times 10$ | 200 |

**Supplementary Table 1 – Measurement parameter of the magnetic island array measured and discussed in Figure 2 of the main text.** Each array, is named here with the same name as in table 1 and figure 2.

**Supplementary Note 1: Sample synthesis and fabrication**

Sample synthesis:

Single crystals of CrGeTe$_3$ were grown using slight excess of Te[1,2]. We grew our samples from high-purity Cr (Alfa Aesar 99.999%), Ge (GoodFellow 99.999%) and Te (GoodFellow 99.999%). Cr, Ge and Te were introduced and sealed in quartz ampoules. Then, the ampoules were heated from room temperature to 930 ºC in 12 h, cooled down to 715ºC in 54 h, and finally cooled down to 500 ºC in 54 h; here the samples remain during 99 h. We quenched the crystals down to ambient temperature by immersion in cold water. We obtained layered crystals of 2 mm × 2 mm × 0.5 mm.

Sample fabrication:

CGT samples were fabricated using the dry transfer technique, which was carried out in a glovebox with an argon atmosphere. The CGT flakes were cleaved using the scotch tape method and exfoliated on commercially available Gelfilm from Gelpack. The CGT flakes were transferred onto a SiO$_2$ substrate. The flakes were exfoliated from the crystals in areas without any Te flux. This was achieved by optically checking that the sample area was free of any inclusions and had large and flat surfaces. The various island shapes were etched or amorphized using a 30 keV Ga$^+$ focused ion beam (FIB)[3].

**Supplementary Note 2: Sample characterization:**

Scanning transmission electron microscopy images of CrGeTe$_3$ islands:

Lamellas were prepared and imaged by Helios Nanolab 460F1 Lite focused ion beam (FIB) - Thermo Fisher Scientific. The site-specific thin lamella was extracted from the CGT array using FIB lift-out techniques[4]. Scanning transmission electron microscopy (STEM) and Energy-Dispersive X-ray Spectroscopy (EDS) analyses were conducted using an Aberration Prob-Corrected S/TEM Themis Z G3 (Thermo Fisher Scientific) operated at 300 KV and equipped with a high-angle annular dark field (HAADF) detector from Fischione Instruments and a Super-X EDS detection system (Thermo Fisher Scientific).

To determine the CGT islands' dimensions, we performed cross-section STEM on all islands presented in Figure 2 of the main text. HAADF STEM images are shown in Supplementary Figure 5.

Energy-Dispersive X-ray Spectroscopy of CrGeTe$_3$ island

In Supplementary Figure 6 we present the High-angle annular dark field (HAADF) image of the $150 \times 150 \times 60$ nm$^3$ island. The image resolves that the crystal structure is damaged due to the FIB etching. Near the etched area, the material is amorphous (bright gray color scale) where the crystalized CGT appears darker. The images reveal the precise thickness of the flake (d = $60 \pm 2$ nm) and the edge cross-section w = $150 \pm 5$ nm. To understand the stoichiometry of the flakes we perform an energy-dispersive spectroscopy (EDS) measurement. The EDS reveals accumulation of Ga and oxidation peaks near the amorphous edge. The concentration decays abruptly over a length of tens nm. We emphasize that Ga concentration peaks appear only in the amorphous part which we found to be non-magnetic. The Ga in the crystalline area less than 2% according to our EDS measurements. Traces of Silicon were also observed in the EDS measurements which seem to originate from organic residues from the PDMS used during the exfoliation process.

Bulk X-ray diffraction
To characterize the bulk crystals, we made x-ray diffraction on crystals milled down to powder (Supplementary Figure 7). X-ray diffraction experiments were performed using an X-ray Bruker D8 Discover Diffractometer at room temperature. Rietveld analysis was performed on X-ray diffractograms using FullProf suite[5]. After Rietveld refinement, we find CrGeTe$_3$ (space group R-3h (148)), with refined lattice parameters a = 0.6809 nm, b = 0.6809 nm, c = 2.0444 nm, with a small trace of Te (see green triangles in Figure).

Bulk SQUID measurements

Bulk magnetic characterization was performed using a Quantum Design SQUID magnetometer. Supplementary Figure 8 depicts the magnetization of a bulk crystal as a function of temperature $M(T)$. We see a clear transition around the expected Currie temperature $T_c = 65$ K. In the inset of Supplementary Figure 8 we show the magnetization as a function of the applied field $M(H)$. At low temperature we obtain the expected value for the saturation magnetization which is around 3 μ$_B$/Cr. This value is consistent with the value obtain in previous reports.[6,7]

**Supplementary Note 3: Uncertainty estimation**

Uncertainty on the island dimensions ($w$ and $d$)

The uncertainties associated with $w$ and $d$ were estimated from the TEM measurements. In samples with $d = 35$ nm, the islands were created by FIB amorphization. Therefore, the uncertainty in $d$ is only due to oxidation, $\delta d = 2$ nm, and with respect to $w$ it is a fraction of the Ga$^+$ beam profile $\delta w = 5$ nm. However, other islands were created by FIB etching, resulting in a trapezoid shape, and the islands are covered by amorphized CGT (which is non-magnetic). In this case, the uncertainty is larger because the amorphous area is not well defined in the TEM image. This results in $\delta d = 10$ to 20 nm, and $\delta w = 10$ to 100 nm, depending on the specific case.

To estimate the uncertainty $\delta p$ in the parameter $p = \frac{w}{V} = \frac{1}{wd}$, we consider the uncertainties associated with each parameter separately and then propagate these uncertainties using the following expression:

$$\delta p = \sqrt{\left(\frac{\partial p}{\partial w} \cdot \delta w\right)^2 + \left(\frac{\partial p}{\partial d} \cdot \delta d\right)^2}$$

Where, $\delta p$, $\delta w$, and $\delta d$ are the the uncertainties corresponding to $p$, $w$, and $d$, respectively.

Uncertainty on the median island coercivity $\widetilde{H_c^i}$

The main source of uncertainty is related to the large transition over which the islands reverse their magnetization. If all island were identical and neglecting thermal fluctuations, all island should reverse their magnetization at the same field $\Delta H_z = H_l - H_f$. In all the measured array, the value of $\Delta H_z$ was rather nearly constant around 73 mT (see Table 1 in the main text). The other source of uncertainty is the field step separating two measurements, which was around $\delta H_z = 2.5$ mT (see supplementary table 1 for the specific values corresponding to each array.

To estimate the resulting uncertainty on the coercive field $\delta H_c$, we consider we sum in quadrature those uncorrelated sources of noise. The uncertainty component generated by $\Delta H_z$ is written as $\frac{H_l - H_f}{2\sqrt{N}}$, where $N$ is the number of island measured in the array (around 100 islands, see Supplementary Table 1 for exact values). In addition given that we consider difference from the mean value we also divide this number by 2. The result is written as follows:

$$\delta H_c = \sqrt{\left(\frac{H_l - H_f}{2\sqrt{N}}\right)^2 + (\delta H_z)^2} \approx 4 \text{ mT}$$

Reproducibility of the coercivity measurement.

To measure the reproducibility in probing the island median coercive field $\widetilde{H_c^i}$, we measure the array's magnetization $M(H_z)/M_{tot}$ for 8 complete hysteresis loops. The results are presented in Supplementary Figure 9a. We calculate the array's coercive field from the data for each of the 16 occurrences. The obtained standard deviation on the value of $H_c$ is 0.7 mT which is smaller than our uncertainty which is dominated by the width of the transition $\Delta H_z / \sqrt{N} = 8$ mT and the steps size in field $\delta H_z = 2.5$ mT. This suggests that the width of the transition $\Delta H_z$ is likely dominated by variability in the island property rather than thermal fluctuations. Further investigation could confirm this finding.

Symmetrization of the $M(H_z)$ curves

We also compare the coercivity measured at $H_z > 0$ and $H_z < 0$ by plotting the $M(H_z)$ for the sweep up with $-M(-H_z)$ for the sweep down (Supplementary Figure 9b). The data overlaps very nicely over the 8 complete hysteresis loops. To further quantify that point, we calculate the array's coercive field for both data sets. We obtain $66.8 \pm 0.4$ mT and $67.8 \pm 0.7$ mT for the sweep down and sweep up data, respectively. This justify the symmetrization of our curves in Figure 2a.

**Supplementary Note 4: Scanning SQUID-On-Tip microscopy**

The SOT was fabricated using self-aligned three-step thermal deposition of Pb at cryogenic temperatures, as described previously[8,9]. The measurements were performed using tips with effective SQUID loop diameters ranging from 145 to 175 nm. All measurements were performed at 4.2 K in a low pressure He of 1 to 10 mbar. The quantification of the measured magnetic field was performed as described previously[8,9].

Reference:

1. Canfield, P. C. Solution Growth of Intermetallic Single Crystals: A Beginner's Guide. in 93–111 (2009). doi:10.1142/9789814261647_0002.
2. Canfield, P. C. & Fisk, Z. Growth of single crystals from metallic fluxes. *Philosophical Magazine B* **65**, 1117–1123 (1992).
3. Noah, A. *et al.* Nano-Patterned Magnetic Edges in CrGeTe3 for Quasi 1-D Spintronic Devices. *ACS Appl Nano Mater* **6**, 8627–8634 (2023).
4. Sezen, M. Focused Ion Beams (FIB) — Novel Methodologies and Recent Applications for Multidisciplinary Sciences. in *Modern Electron Microscopy in Physical and Life Sciences* (InTech, 2016). doi:10.5772/61634.
5. Rodríguez-Carvajal, J. Recent advances in magnetic structure determination by neutron powder diffraction. *Physica B Condens Matter* **192**, 55–69 (1993).
6. Carteaux, V., Brunet, D., Ouvrard, G. & Andre, G. Crystallographic, magnetic and electronic structures of a new layered ferromagnetic compound Cr2Ge2Te6. *Journal of Physics: Condensed Matter* **7**, 69–87 (1995).
7. Ji, H. *et al.* A ferromagnetic insulating substrate for the epitaxial growth of topological insulators. *J Appl Phys* **114**, (2013).
8. Vasyukov, D. *et al.* A scanning superconducting quantum interference device with single electron spin sensitivity. *Nat Nanotechnol* **8**, 639–644 (2013).
9. Anahory, Y. *et al.* SQUID-on-tip with single-electron spin sensitivity for high-field and ultra-low temperature nanomagnetic imaging. *Nanoscale* **12**, 3174–3182 (2020).